# Measuring Motional Dynamics of $(CH_3)_2NH_2^+$ in the Perovskite–like Metal-Organic Framework $[(CH_3)_2NH_2][Zn(HCOO)_3]$: The Value of Low-Frequency EPR

*Sylvain Bertaina[1]\*, Nandita Abhyankar[2], Maylis Orio[3], Naresh S. Dalal[2,4]\**

1. Aix-Marseille Université, CNRS, IM2NP UMR 7334, 13397 Marseille Cedex 20, France.

2. Department of Chemistry & Biochemistry, Florida State University, 95 Chieftain Way, Tallahassee Florida 32306, United States.

3. Aix Marseille Université, CNRS, Cent. Marseille, iSm2, 13397 Marseille Cedex 20, France.

4. National High Magnetic Field Laboratory (NHMFL), 1800 East Paul Dirac Drive, Tallahassee, Florida 32310, United States.

**Corresponding Authors**

sylvain.bertana@im2np.fr

dalal@chem.fsu.edu

ABSTRACT: Dimethylammonium zinc formate (DMAZnF) is the precursor for a large family of multiferroics, materials which display co-existing magnetic and dielectric ordering. However, the mechanism underlying these orderings remains unclear. While it is generally believed that the dielectric transition is related to the freezing of the order-disorder dynamics of the dimethylammonium (DMA$^+$) cation, no quantitative data on this motion are available. We surmise that this is due to the fact that the timescale of this cationic motion is on the borderline of the timescales of experimental techniques used in earlier reports. Using multifrequency EPR, we find that the timescale this motion is ~ 5 x 10 $^{-9}$ s. Thus, S-band (4 GHz) EPR spectroscopy is presented as the technique of choice for studying these motional dynamics. This work highlights the value of the lower-frequency end of EPR spectroscopy. The data are interpreted using DFT calculations and provide direct evidence for the motional freezing model of the ferroelectric transition in these metal-organic frameworks with the ABX$_3$ perovskite-like architecture.

## 1 - INTRODUCTION

During the last decade or so, there has been considerable research interest in the family of metal-organic framework (MOF) compounds with the general perovskite formula ABX$_3$, because of their structural versatility and potential applications[1–6]. A particularly popular class of such MOFs consists of formate frameworks[5–9] with the general formula [A][M(HCOO)$_3$]. Here, A$^+$ is a molecular cation such as the dimethylammonium ([(CH$_3$)$_2$NH$_2$$^+$]) or DMA$^+$ cation, and M$^{2+}$ is a 3d transition metal cation. Like their inorganic counterparts, such as SrTiO$_3$, these MOFs exhibit two primary, spontaneous, and switchable internal domain alignments in the same phase: the alignment of electric polarization, as seen in ferroelectrics, and the alignment of magnetization, as seen in ferromagnets. More importantly, their electric polarization can be reversibly switched by an external magnetic field and the magnetization altered reversibly by

an external electric field[1]. It has been emphasized in earlier reviews[1–4] that in order to optimize their device potential, it is important to obtain a deeper understanding of the molecular interactions that underlie their ferroelectric and ferro(antiferro)magnetic phase transitions. In this regard, it is known that for these MOFs, the motional dynamics of the $A^+$ cation seem to play a pivotal role in driving the structural, ferroelectric transition[6–10]. In the high-temperature, paraelectric phase, the $A^+$ molecular unit, which possesses an axis of three-fold symmetry, appears to be dynamically disordered over three possible orientations[5–8]. The phase transition coincides with the loss of this three-fold symmetry, as suggested by X-ray diffraction[5–7], neutron scattering[11], and IR[12], Raman[12,13], EPR[14–17] and NMR[18,19] spectroscopies. However, none of these techniques has provided a quantitative estimate of the correlation time or frequency of motion of the $DMA^+$ cation, a much sought-after datum in the phase transition model of these MOFs.

This report presents a variable frequency EPR study aimed at measuring the timescale of motional dynamics of the $A^+$ cation $[(CH_3)_2NH_2]^+$ in dimethylammonium zinc formate, $[(CH_3)_2NH_2]Zn(HCOO)_3$, henceforth referred to as DMAZnF. This sample was chosen because in this family of MOFs,[7,8] DMAZnF was the first compound shown to exhibit a dielectric transition. It has now been confirmed to be a ferroelectric below the dielectric phase transition[8,9,20].

DMAZnF displays a first-order structural phase transition at about $T_C$=170K. X-ray diffraction reveals that at high temperature (HT), the compound exists in the trigonal R3c phase while at low temperature (LT), its structure belongs to the monoclinic noncentrosymmetric Cc space group[10]. Despite this structural phase transition, the global structure does not change significantly. Figure 1 shows the crystal structure of DMAZnF in the HT phase. The $Zn(HCOO)_3^-$ subsystem, in which the $Zn^{2+}$ ions are connected through the formate groups,

forms a cage in which the DMA$^+$ cations are located. The symmetry of the HT phase allows the nitrogen atom to occupy one of three equivalent positions. In Figure 1, we have differentiated these positions by three colors: yellow, green, and blue. The nitrogen continuously moves between these three positions (as represented by the round arrow). The position of any DMA$^+$ ion is independent of the positions of the neighboring DMA$^+$ ions in the paraelastic/paraelectric phase. The symmetry of the LT phase (Figure SI1) allows only one position for the N atom. In the LT ferroelastic/ferroelectric phase, DMA$^+$ cations are ordered in a checkerboard arrangement, with alternate DMA$^+$ cations pointing in slightly different directions.

Electron paramagnetic resonance (EPR) spectroscopy is the tool of choice to probe the anisotropy and crystal field effects acting on paramagnetic ions. In particular cases, it is also possible to probe the fluctuation of the crystal field in time. In this paper, we focus only on the HT phase (T>170K), in which the DMA$^+$ cations are thought to undergo hindered rotational reorientations between three equivalent sites within the zinc formate cage. In order to perform EPR experiments, we have doped the DMAZnF (diamagnetic) with a small (~ 0.1 mole %) amount of Mn$^{2+}$, which acts as a paramagnetic probe of crystal-field fluctuations due to the motion of the DMA$^+$ cations around it. Thus, the timescale of the motion of DMA$^+$ can be measured via the Mn$^{2+}$ spin probe.

We show that the lack of measurements of the timescale of motional dynamics of DMA$^+$ can be ascribed to the fact that this motional timescale falls well below the timescales of optical-IR spectroscopy and neutron scattering, which are techniques used in earlier studies of these materials. As detailed below, this timescale falls in the range of the rarely used low-frequency (~ 4 GHz) S-band EPR technique. The obtained data, supplemented by theoretical modeling (DFT calculations), provide strong support for the dynamic behavior of the DMA$^+$ cation and its role in the ferroelectric transition.

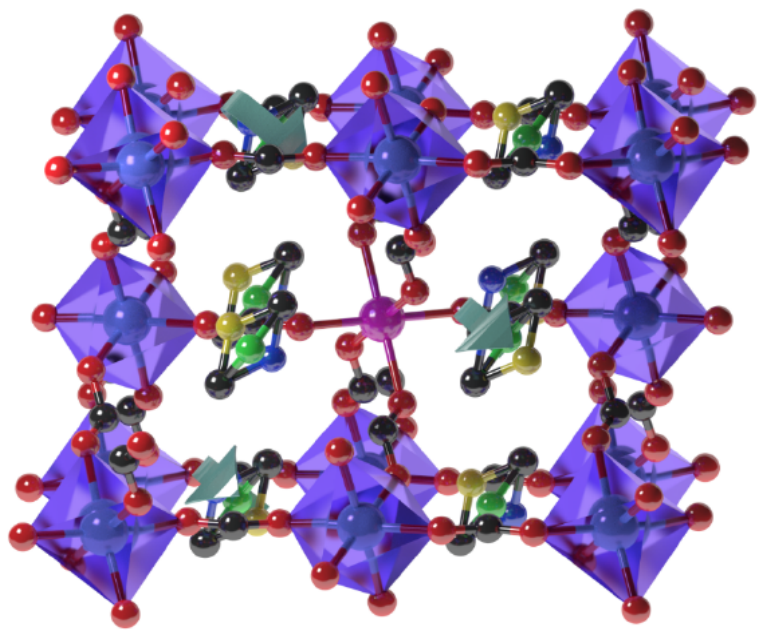

Figure 1: Schematic representation of $Mn^{2+}$-doped DMAZnF in the HT phase. The nitrogen atom in $DMA^+$ can occupy one of three equivalent positions, as represented by three different colors (dark blue, yellow, and green). The reorientation of $DMA^+$ is thermally activated, resulting in hindered rotation between these three positions. To probe the effect of this motion by EPR spectroscopy, $Zn^{2+}$ (light blue) was replaced by $Mn^{2+}$ (pink) at a nominal concentration of 0.1%. Oxygen atoms are shown in red and carbon atoms are shown in black. Hydrogens have been omitted to simplify the presentation. The structure of the LT phase is presented in Fig SI1.

## 2- EXPERIMENTAL DETAILS:

Synthesis: Single crystals of DMAZnF were grown by solvothermal methods reported previously. A reaction mixture containing 1 mmol anhydrous $ZnCl_2$, 6 mL DMF, and 6 mL deionized $H_2O$ was spiked with trace amounts of $MnCl_2$. The exact amount of Mn in the final product was not determined but by comparison with the EPR-standard free radical

diphenylpicrylhydrazyl (DPPH), the amount was assessed as less than 1 mole % in the final product. The mixture was heated overnight in a Teflon-lined autoclave. The solution was gradually cooled over a few hours, and the supernatant was set aside to crystallize. Over a period of a few days, block-like single crystals with dimensions of 1-3 mm were obtained.

Single crystal diffraction: Single crystal X-ray diffraction was employed to confirm the crystal structure and the relationship between the external morphology and unit-cell and crystal axes. Figure 1 shows a schematic of the pseudocubic crystal structure of DMAZnF, with eight Zn ions at the corners of the cube and the $DMA^+$ cation at the center. As is known, DMAZnF undergoes a paraelectric-to-ferroelectric phase transition at $T_C$ = 166 K, with a hysteresis of less than 1 K, in agreement with our heat capacity studies[7]. In the currently accepted model of the ferroelectric transition, at $T > T_C$, the $DMA^+$ cations are dynamically disordered between three positions, with $1/3^{rd}$ site occupation at each position, as depicted in Fig. 1. At $T_C$, the dynamic disordering ceases and the $DMA^+$ cations become ordered. As mentioned in the Introduction, despite several studies using neutron scattering and IR, Raman, EPR and NMR spectroscopies, it has not been possible to pin down the actual timescale or frequency of the order-disorder dynamics of the $DMA^+$ cation. This issue provided the main motivation for the present undertaking.

EPR spectroscopy:

EPR experiments were performed using three conventional Bruker spectrometers operating at S-band (E600 - 4 GHz), X-band (EMX - 9.6 GHz), and Q-band (E600 - 34 GHz), respectively. The sample was mounted on a suprasil rod and its orientation was measured using a goniometer. All measurements were performed with H//[102] axis. The temperature was varied from 300K to 170K (the phase transition temperature), using an Oxford ITC temperature

controller. 10 min were allowed for stabilization at each temperature. Magnetic field modulation associated with lock-in detection was employed, resulting in the derivative of the signal. All measurements were carried out with the static field H perpendicular to the (012) face of the crystal. It is noted that the sensitivity of the S-band spectrometer is about an order of magnitude lower compared to those at X and Q-band. Therefore, the signal-to-noise ratio is correspondingly worse in the S-band spectra.

## 3- COMPUTATIONAL DETAILS

All theoretical calculations were based on the Density Functional Theory (DFT) and were performed with the ORCA program package[21]. To facilitate comparisons between theory and experiments, all DFT models were obtained from the experimental X-ray crystal structures, and were optimized while constraining the positions of all heavy atoms to their experimentally derived coordinates. Only the positions of the hydrogen atoms were relaxed because these are not reliably determined from the X-ray structure. Geometry optimizations were undertaken using the GGA functional BP86[22–24] in combination with the TZV/P[25] basis set for all atoms and by exploiting the resolution of the identity (RI) approximation in the Split-RI-J variant[26] with the appropriate Coulomb fitting sets[27]. Increased integration grids (Grid4 and GridX4 in ORCA convention) and tight SCF convergence criteria were used. The zero-field splitting (ZFS) parameters were obtained from single-point calculations using the BP functional. Scalar relativistic effects accounted for the metal center, and were included with ZORA paired with the SARC def2-TZVP(-f) basis sets[28] and the decontracted def2-TZVP/J Coulomb fitting basis sets for all atoms. Increased integration grids (Grid4 and GridX4 in the ORCA convention) and tight SCF convergence criteria were used in the calculation. The spin-spin contribution to ZFS was calculated on the basis of the UNO determinant[29].

## RESULTS AND DISCUSSION

<u>In the rigid model:</u>

EPR spectra of DMAZnF:Mn²⁺ have been reported recently,[16] with particular attention paid to the LT phase. However, the dynamics of interest occur in the HT phase. We therefore focused mainly on the dynamics in the HT phase.

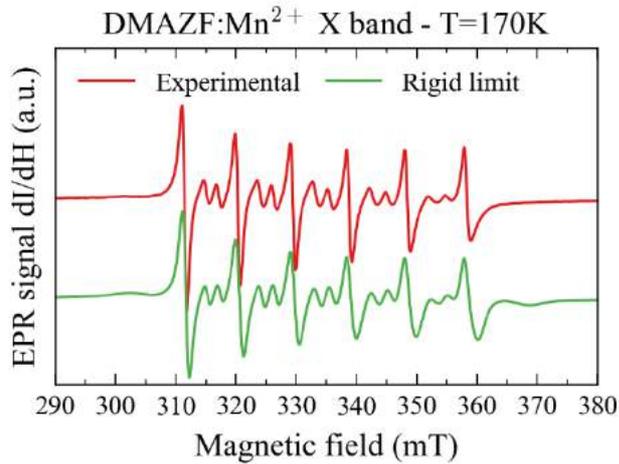

Figure 2: X-band EPR spectra of the $Mn^{2+}$ spin probe in DMAZnF, recorded just above Tc. The experimental data (red) are fitted (green) using the parameters provided in the text.

Figure 2 shows an example of a spectrum recorded at X-band and T=170K (just above the ferroelectric phase transition temperature). In general, the spectrum of a $Mn^{2+}$ ion (S=5/2, I=5/2) is composed of six sets of 5 allowed transitions ($\Delta m_S=\pm 1$, $\Delta m_I =0$), giving 30 allowed lines. In our case, the spectrum is composed of only 6 allowed $\Delta m_S=\pm 1$ ($m_I$=-1/2 to 1/2) transitions. The ($m_S=\pm 5/2$ to $\pm 3/2$, and $m_S=\pm 3/2$ to $\pm 1/2$) transitions are invisible because while they do occur, they are too broad to be resolved. Between the allowed transitions, we observed 10 weaker forbidden ($\Delta m_S=\pm 1$, $\Delta m_I=\pm 1$) transitions[30]. The presence of the forbidden transitions proves that there is a non-negligible crystal field anisotropy. Multiple-quantum transitions, observed in weak anisotropy,[31,32] are not seen here.

To simulate the EPR spectra, we used the following Hamiltonian:

$$\mathcal{H} = g\mu_b \vec{H}\vec{S} + A\vec{S}\vec{I} + \sum_k \sum_q B_k^q \hat{O}_k^q(\vec{S})$$

Here, the first term represents the Zeeman interaction, the second the hyperfine interaction, while the last term represents the crystal field interaction.

In solid-state EPR spectroscopy, the signal is usually described by the rigid model, in which the motion of atoms is small compared to the timescale ($\sim 1/f$, f = frequency) of measurement. We used the Matlab module Easyspin[33] to simulate the spectra.

The green line shows the spectrum simulated using the standard rigid model and the following parameters: g=2.01, A=-265MHz and D=$B_0^2$/3=250($\pm$50)MHz. To take into account the large anisotropy strain, we used a Gaussian distribution of D, $\Delta$D~150MHz. The simulation describes the experimental data accurately, and the fit parameters are in agreement with those reported in other EPR studies[15]. Due to the very large line broadening, it was not possible to extract other crystal field parameters from the data.

We are conscious that the physical meaning of this very large value of D-strain is not obvious, but we present a plausible hypothesis based on our theoretical calculations (vide infra).

Compared to the nominal value of D=250MHz, the value of $\Delta$D is not at first understandable since $\Delta$D, which is caused by small local strains in the lattice, should be a perturbation. Here we found a strain of about 60%, which should destroy the crystal structure. However, another scenario can explain this large value. We recall that the $Mn^{2+}$ spin is surrounded by eight nearest-neighbor $DMA^+$ cations. In the HT phase, each $DMA^+$ can occupy one of three equivalent positions, and is constantly moving between these positions. However, if the timescale of the measurement is fast enough (faster than the time of motion between two positions), the $DMA^+$ cations appear frozen in one configuration. We can estimate the number of configurations of eight $DMA^+$ cations surrounding one $Mn^{2+}$ spin. With eight $DMA^+$ cations, each capable of occupying one of three equivalent positions, we have $3^8 = 6561$ possible configurations. The sample contains many billions of $Mn^{2+}$ spins, each experiencing a crystal field corresponding to one of these 6561 configurations. Thus, the EPR signal recorded is like a "snapshot" of all the configurations at the same time. All configurations are equally probable and each gives a different EPR spectrum. The spectrum shown in Fig. 2 is actually the sum of all 6561 spectra. We simulated this effect using a Gaussian distribution of D.

In order to confirm this model, we used DFT. We have estimated the ZFS of selected configurations. We chose to work with a minimal model consisting of one $Mn^{2+}$ ion bound to six formate anions and surrounded by eight $DMA^+$ cations (see figure SI2). The resulting metal cluster displays a quasi-octahedral coordination geometry. Based on the HT X-ray structure that identified three equivalent positions of the nitrogen atom in each $DMA^+$, we considered several configurations in which the Mn-N distances can vary from 4.61 to 5.78 Angstroms. This provided a random sampling of the possible configurations (see Table SI3). The distribution of the ZFS parameter was found to be $\Delta D_{DFT} \sim 125$ MHz, in rather good agreement with the experimentally estimated $\Delta D \sim 150$ MHz .

 The results of DFT calculations lead to several conclusions. Firstly, the crystal field effect on the $Mn^{2+}$ spin has previously been considered to originate solely from the surrounding octahedral oxygens. Here we prove that the $DMA^+$ cations also play an important role in determining the crystal field of the $M^{2+}$ ion. The reason is that although the oxygens are in first-neighbor positions, they form an octahedron with a nearly cubic symmetry and consequently, a weak anisotropy. On the contrary, although the $DMA^+$ cations are farther away than the oxygens, they create a low-symmetry crystal field around $Mn^{2+}$.

Secondly, in the HT phase, the $DMA^+$ cations are considered to be undergoing hindered rotation between their three equivalent positions[18,19] but if the timescale of the measurement is small enough, the cation appears frozen. This is the rigid limit, which is the case for X-band measurements at low temperature (170K<T<200K) and Q-band measurements at all temperatures.

<u>In the slow-motion regime:</u>

Intriguingly, at higher temperatures (>200K) in X-band measurements, the lineshape is different compared to the one in Fig. 2.

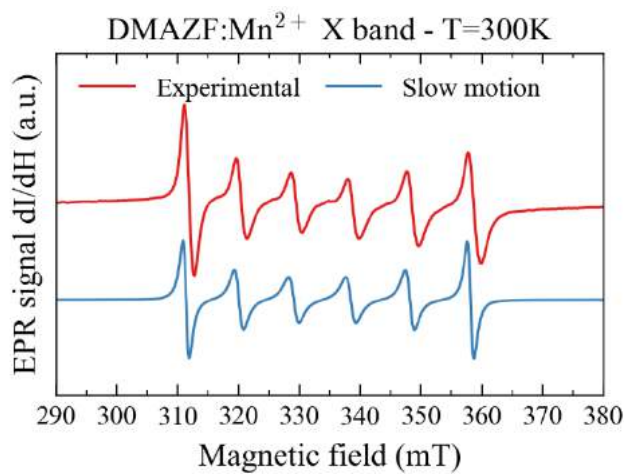

Figure 3: X-band EPR spectra of the $Mn^{2+}$ spin probe in DMAZnF, recorded at room temperature. The experimental data (red) are compared to the EPR signal expected in the slow-motion regime (blue).

Figure 3 shows the X-band EPR spectrum of DMAZF:$Mn^{2+}$ at room temperature. The forbidden transitions disappear and the linewidths are smaller for the outer lines ($m_I=\pm5/2$) than for the central lines ($m_I=\pm1/2$). The rigid-model simulation did not yield the experimental lineshape at room temperature, for any chosen values of fit parameters. This is because the rigid limit assumes that the crystal field is constant in time (i.e. the atoms are fixed) and the linewidth is included phenomenologically, as a distribution of crystal fields, in the Hamiltonian. On the contrary, in the HT phase, the DMA$^+$ cation can occupy one of three equivalent positions and is almost free to move between these positions. Depending on the rate, the motion of DMA$^+$ must be taken into account and the rigid limit is no longer valid. The snapshot described before is now blurred by the motion of the cation, which is comparable or fast compared to the timescale of the measurement. This model is called the "slow-motion" model and has been used in the past to describe the EPR spectra of viscous solutions[34–36]. In these papers, the Mn$^{2+}$ ions were moving in a viscous solution, causing fluctuation of the anisotropic interaction with the solvent as a function of time. In our case, the Mn$^{2+}$ ion is fixed but the motion of the DMA$^+$ cations leads to the same effect as that observed for Mn$^{2+}$ spins surrounded by a viscous solution.

As is known[34–36], the choice of model is based on the correlation time ($\tau_c$) of the process modulating the ZFS. Depending on the value of $\tau_c$ relative to the timescale of measurement (inverse of the microwave frequency), the DMA$^+$ is seen as either rigid or moving. If $\tau_c < 1/f$, then during the timescale of the measurement, the DMA$^+$ appears to be moving between the three possible positions: this is the slow-motion regime. If $\tau_c > 1/f$, then the DMA$^+$ appears to be frozen in one of the three positions during the measurement: this is the rigid limit mentioned above. Figure 3 shows the simulation in the slow-motion regime, using D=250MHz, A=265MHz and $\tau_c$ =2.10$^{-9}$ s. This model describes the room-temperature spectrum accurately. In order to support our model, we acquired EPR spectra of DMAZF:Mn$^{2+}$ at a lower frequency (S-band – 4GHz), where the effect of motional broadening is expected to be enhanced. Unfortunately, the sensitivity of the S-band spectrometer is about 2 orders of magnitude lower than at X-band. As a consequence, the signal-to-noise ratio of the measurements is poor and the signals are noisy.

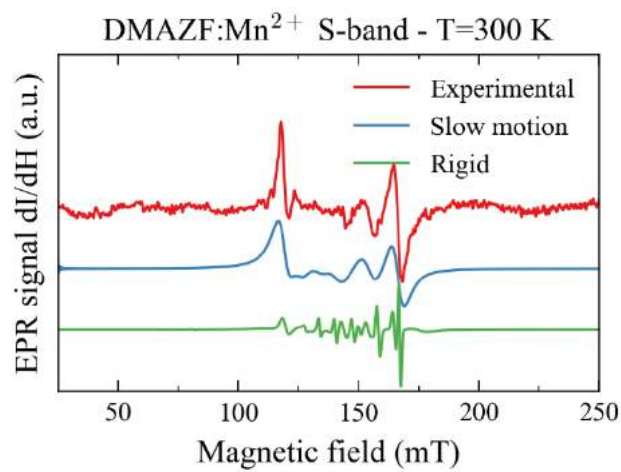

Figure 4: S-band (4 GHz) EPR spectra of the $Mn^{2+}$ spin probe in DMAZnF, recorded at room temperature. The experimental data (red) are compared to the slow-motion regime (blue) and the rigid model (green). Good agreement with the slow-motion regime is clearly evident.

Figure 4 shows a spectrum recorded at S-band at room temperature. The purpose of this measurement was to compare the two models describing the X-band results. The blue line is the simulated spectrum corresponding to the slow-motion regime, obtained using the correlation time extracted from X-band results (Fig. 3). The green line is the simulated spectrum corresponding to the rigid limit, obtained using the Hamiltonian parameters extracted from Fig. 2 (g=2.01, A=-265MHz and D=$B_0^2$/3=250MHz and ΔD~150MHz). Although the linewidth is slightly overestimated, the slow-motion regime describes the experimental data much more accurately than the rigid limit.

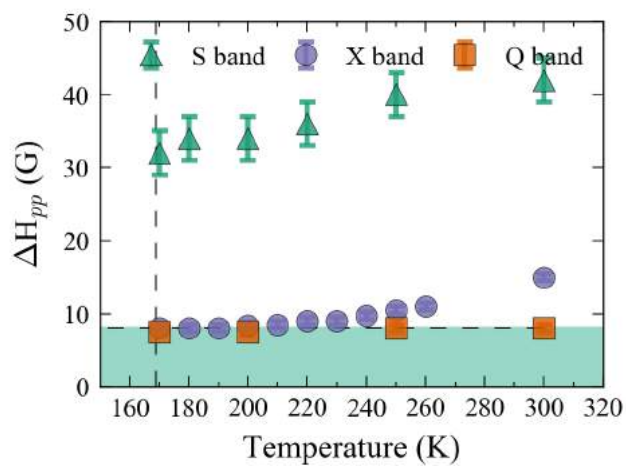

Figure 5: Peak-to-peak linewidth of the rightmost EPR line, recorded between 300K and 170K for the three frequencies used. The vertical dashed line is the structural transition temperature. The horizontal dashed line is the inhomogeneous linewidth.

Finally, Fig. 5 compares the linewidths for the three frequencies as a function of temperature. At Q-band, we find the linewidth of 8 G to be nearly independent of temperature. At X-band, the linewidth starts at 15 G at room temperature and then decreases with temperature, finally reaching a minimal value of about 8 G. At S-band, the linewidth is about 40 G, and decreases with temperature to 30 G. Applying the slow-motion model[36] with $\tau_c = 2.10^{-9}$ s ( at T=300K), we obtained linewidths of 46 G at S-band, 15 G at X-band, and 1 G at Q-band. While the simulated linewidths at S-band and X-band are in good agreement with experimental results, those at Q-band are not. This is because the timescale of Q-band measurements is very fast compared to the correlation time of the motion of the DMA$^+$ ion. Thus, the system is best described by the rigid-limit model, in which inhomogeneous broadening of the linewidth has been estimated to be 8 G. This value is the threshold below which slow-motion broadening is hidden by the inhomogeneous broadening discussed previously. At X-band, both regimes occur. At high temperature, slow-motion broadening is dominant. Upon decreasing the temperature, the correlation time increases and below about 220K, the DMA$^+$ cation appears frozen at the measurement timescale and the rigid limit is reached.

The timescale of spectroscopy is an important parameter to take into account, especially if the motion inside the system is important. This is the case in the DMA formate family, in which the nature of the multiferroicity is directly connected to the motion of DMA$^+$. In high-energy spectroscopy (X-ray, neutron), the timescale is extremely short and the DMA$^+$ ions appear frozen in the HT phase, giving the reported result of three equivalent positions of nitrogen[7]. On the other side, in low-energy spectroscopy, the DMA$^+$ appears to be moving, irrespective of the experimental temperature[19]. Dielectric susceptibility measurments[18] have shown that the motion of the dipoles, which is directly related to the motion of the DMA$^+$ ions, has a characteristic frequency $f_0=2.97 \times 10^{13}$ Hz and energy $E_a= 3600$K (0.31 eV) . These values lead to a correlation time of the dipole $\tau_c = 1/f_0 \exp(E_a/kT)$, of $5.10^{-9}$ s at T=300K. This is in reasonably good agreement with the value of the EPR correlation time of $2.10^{-9}$s. It also agrees with the correlation time obtained by a recent $^{15}$N NMR study[37].

CONCLUSION:

This work extends earlier EPR studies of the phase transition of the metal-organic framework compound DMAZnF, using the Mn$^{2+}$ (S = 5/2) paramagnetic ion as a spin probe. Emphasis is laid on understanding the relative signal intensities of the hyperfine peaks in the HT, paraelectric phase of this compound; which is the parent of this class of MOFs, as detailed in the Introduction. In particular, Simenas and coworkers[17] have used pulsed EPR , ENDOR, and CW EPR to probe the motion of the DMA$^+$ cation , but these studies focused more on the order parameter than on the large spread in the value of the D parameter in the paraelectric phase. The authors did perform an Arrhenius analysis of the motional effect based on temperature variation of the linewidth, but a quantitative analysis of the rate of motion was not described.

We argued that the timescale of the motion is slow enough that going to frequencies lower than X-band might provide fruitful information about this motion. Indeed, S-Band (4 GHz) EPR spectra yielded clear signatures of the slow motion of both the formate and the $DMA^+$ groups; the rate of hindered rotation of $DMA^+$ is well described by the correlation time of $2.10^{-9}$ s at ambient temperature. Additionally, the unusual result that $\Delta D$, i.e. the spread in D, is larger than D itself can be rationalized by our DFT calculations. We anticipate that our studies will spark new interest in lower-frequency EPR , which can be combined with DFT studies to understand the role of molecular motion in the mechanism of the ferroelectric phase transition in these hybrid MOFs.

SUPPORTING INFORMATIONS:

Crystallographic structure of the LT phase. Minimal model used for the DFT calculation and the results of the 14 configurations used.

ACKNOWLEDGEMENTS

 We thank CNRS' research infrastructure RENARD (FR3443) for EPR facilities, and the US National Science Foundation, via Award CHE-1464955.  The NHMFL is supported by the Co-operative Agreement Grant No. DMR-1157490 and the State of Florida.

TOC Graphic

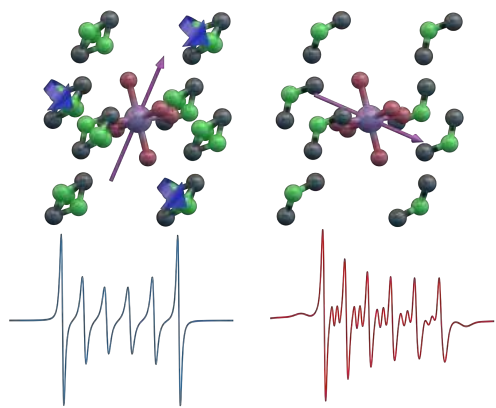

TOC : Effect of the motion of the DMA cation on the EPR spectrum of the $Mn^{2+}$ probe. Left (right) : the motion is fast (slow) compared to the time scale of the measurement.